\documentclass[fleqn,10pt]{wlscirep}
\title{Materials analysis and focused ion beam nanofabrication of topological insulator Bi$_2$Se$_3$}

\author[1,+]{Sarah Friedensen}
\author[1,+]{Jerome T. Mlack}
\author[1,*]{Marija Drndi\'{c}}

\affil[1]{Department of Physics and Astronomy, University of Pennsylvania, Philadelphia, Pennsylvania 19104, USA}

\affil[*]{drndic@physics.upenn.edu}

\affil[+]{these authors contributed equally to this work}

\begin{abstract}
Focused ion beam milling allows manipulation of the shape and size of nanostructures to create geometries potentially useful for opto-electronics, thermoelectrics, and quantum computing. We focus on using the ion beam to control the thickness of Bi$_2$Se$_3$ and to create nanowires from larger structures. Changes in the material structure of Bi$_2$Se$_3$ nanomaterials that have been milled using a focused ion beam are presented. In order to characterize the effects of ion beam processing on the samples, we use a variety of techniques including analytical transmission electron microscopy and atomic force microscopy. The results show that while part of the material remains intact after shaping, amorphous regions form where the beam has been used to thin the sample. For wires created by thinning the material down to the substrate, the sidewalls of the wires appear intact based on diffraction images from samples cut at an angle, but thin crystalline regions remain at the wire edges. Even with the resulting defects, focused ion beam milling shows promise for directly fabricating intricate nanodevices of Bi$_2$Se$_3$.
\end{abstract}
\begin{document}

\flushbottom
\maketitle
%
%
\thispagestyle{empty}

\section*{Introduction}
In the field of nanoelectronics some of the most interesting device proposals require the control of nanostructure properties, especially the shape and size. In the case of nanodevices made from topological insulators, such as bismuth selenide (Bi$_2$Se$_3$), proposals exist for observing exotic phenomena that could have far-reaching applications, including fault-tolerant quantum computing using Majorana fermions\cite{Ilan_2014,Fu_Kane_2008,AR_Akhmerov_2009,T_Hsieh_2012,Beenakker_Review,Kitaev,Nayak} and dramatic increases in thermoelectric efficiency\cite{GLSun_2015,Tretiakov_2012,Tretiakov_2011,Shi_2015,HSShin_2016,KKWu_2016}. These proposals require the ability to fabricate structures such as wires with well-controlled dimensions\cite{Ilan_2014,HSShin_2016}, antidot lattices with specific spacings\cite{Tretiakov_2012,Tretiakov_2011}, and carefully patterned defects\cite{Shi_2015,Sacksteder_2015,Philip_2017}. Directly growing nanotstructures bottom-up into the necessary geometries and defect densities would be ideal, but controlled growth and scaling of such devices is complicated. An alternative is to use top-down methods, such as ion milling, to modify the Bi$_2$Se$_3$\cite{Bhattacharyya_QOsc,Bhattacharyya_SdH,PASharma_2014,WZhou_2017,Tan2016,Fukui_2014,ASharma_2016,Childres_2013,Saji_2005,Jia2016}.  

\begin{figure*}[ht]
\centering
\includegraphics{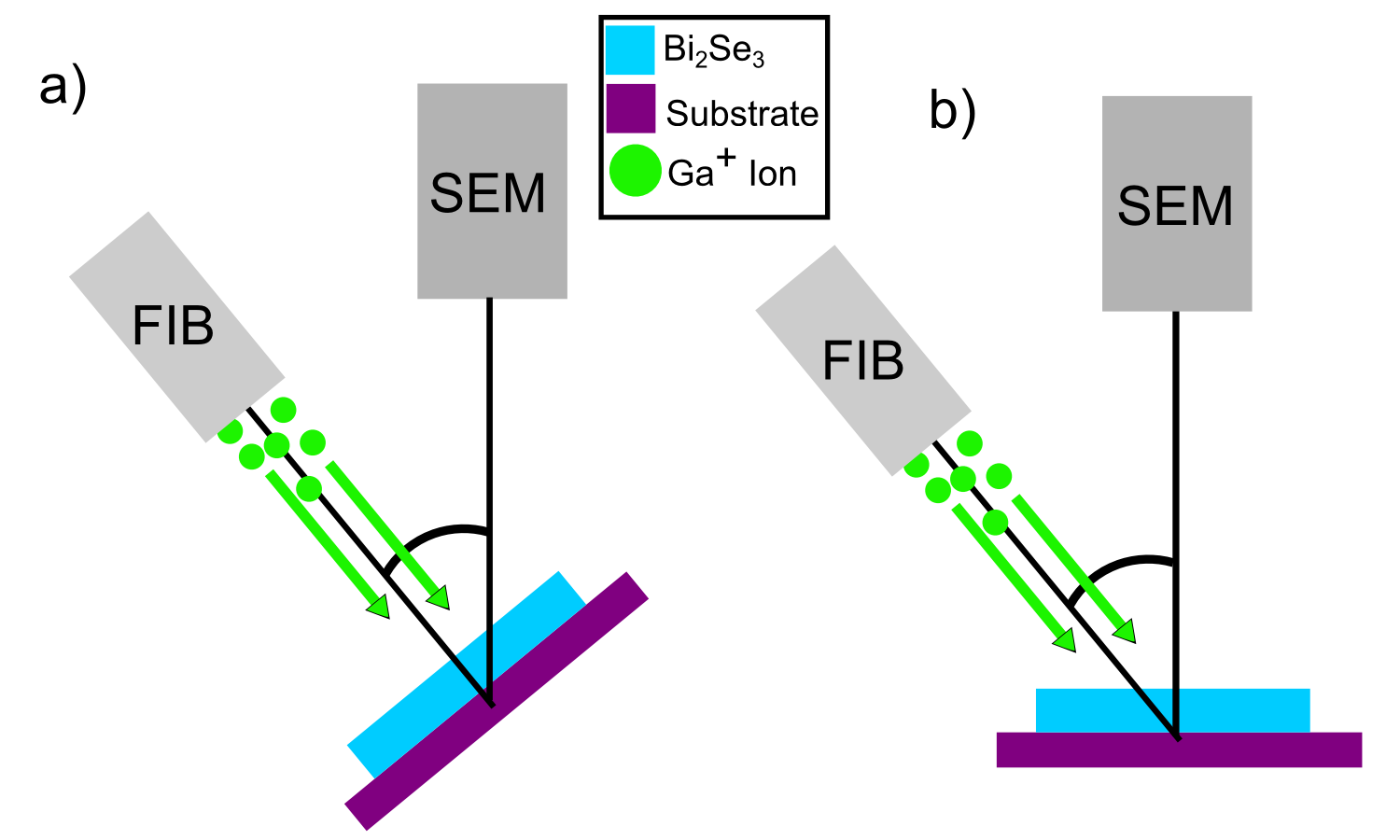}
\caption{\label{diagram}
Diagram of FIB milling setup. (a) With Bi$_{2}$Se$_{3}$ flake aimed toward FIB beam for straight-on milling. (b) With Bi$_{2}$Se$_{3}$ flake angled away from FIB beam for angled milling.
}
\end{figure*}

In this article, we focus on the use of gallium-based focused ion beam (FIB) milling, diagramed in Fig. \ref{diagram}. FIB milling utilizes Ga$^{+}$ ions to sputter away a target material and can be used to both sculpt the material\cite{Bhattacharyya_SdH,Bhattacharyya_QOsc,ASharma_2016,Fukui_2014} and introduce defects\cite{Fukui_2014,PASharma_2014}. Recent transport studies have shown that FIB-created nanowires of Bi$_2$Se$_3$ exhibit increased photoconductivity at room temperature\cite{ASharma_2016} and retain signatures of their topological surface states at low temperature\cite{Bhattacharyya_SdH,Bhattacharyya_QOsc}. The low-temperature transport results indicate, however, that the nanowires are not entirely Bi$_2$Se$_3$ and that the structure has been partially altered by the FIB\cite{Bhattacharyya_QOsc}. While these results are encouraging for FIB milling as a method of top-down fabrication of Bi$_2$Se$_3$, they do not fully address the materials analysis and changes in the material structure. Therefore, in order to better understand the effects of FIB imaging and modification on Bi$_2$Se$_3$, we have fabricated nanostructures via thinning and cutting with the ion beam and analyzed the material changes. In our study, we combine transmission electron microscopy (TEM) methods---such as energy dispersive x-ray spectroscopy (EDS) and selected area diffraction (SAED)---with atomic force microscopy (AFM).

\section*{Results}
\begin{figure*}[ht]
\centering
\includegraphics{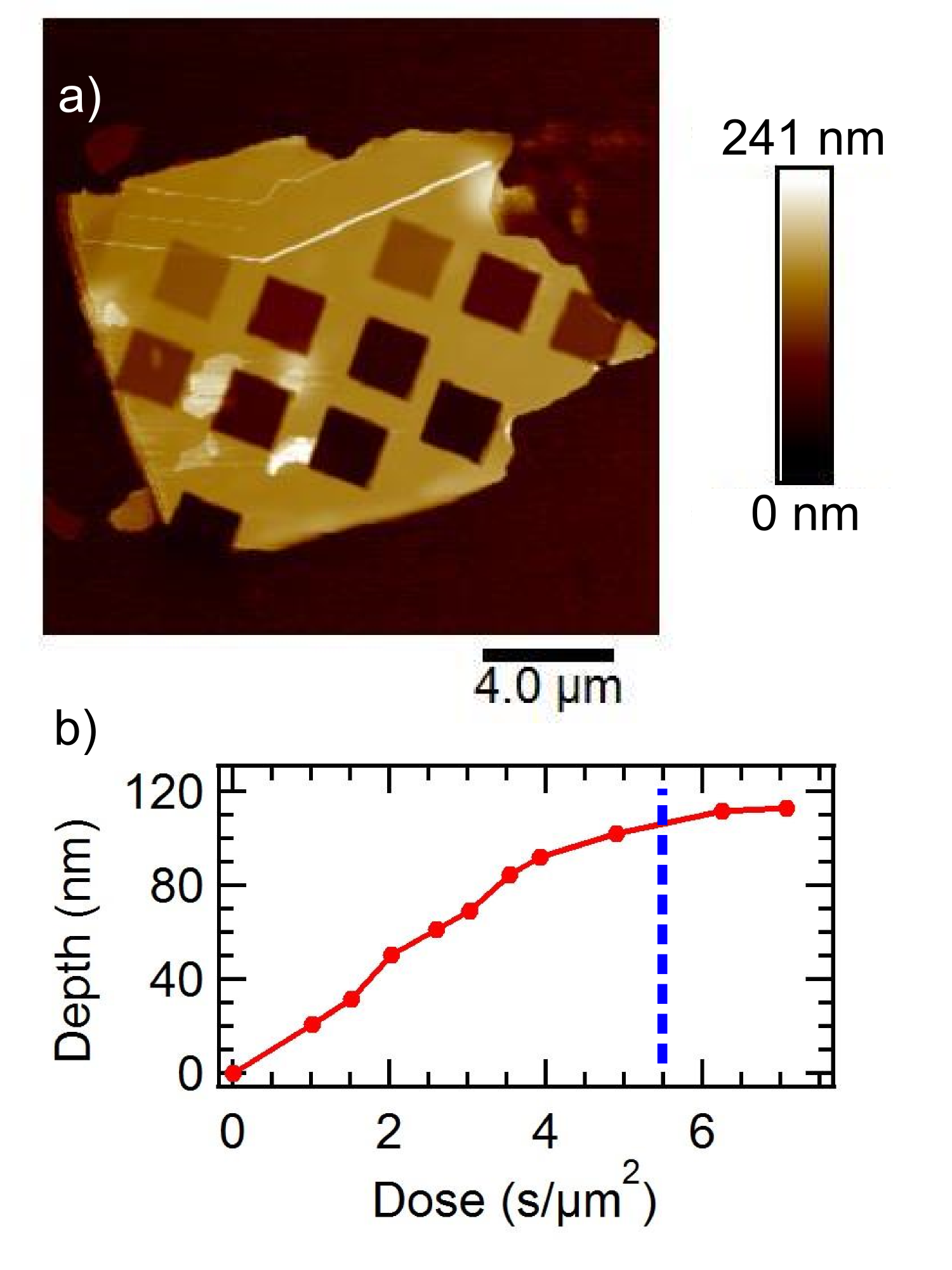}
\caption{\label{Dose_Test}
Bi$_{2}$Se$_{3}$ flake thinned using focused ion beam at 10 pA setting. (a) AFM image of FIB thinned Bi$_{2}$Se$_{3}$ on silicon oxide substrate. Each square is approximately $\mu$m$^2$ in area. (b) Plot of depth as a function of FIB beam dose. Dashed blue line represents the approximate dose at which the beam has milled through the Bi$_{2}$Se$_{3}$ and begun milling the silicon oxide. 
}
\end{figure*}

An AFM map of a FIB-milled Bi$_2$Se$_3$ flake on a SiO$_2$ substrate is shown Fig. \ref{Dose_Test}(a). Multiple 4 $\mu$m$^2$ regions of the flake were exposed to the ion beam, each for a different length of time. The shortest mill time was 4 seconds, and the longest was 28 seconds. All milled regions have a roughness of less than 1 nm, which indicates uniform milling across each area. In comparison, Argon plasma etching\cite{Childres_2013} creates surfaces with roughness greater than 2 nm for any etch time over a few seconds. Fig. \ref{Dose_Test}(b), which shows the dose in units of seconds per unit area versus etch depth, was calculated from the map in Fig.\ref{Dose_Test}(a). The mill-rate appears constant until a depth of 110 nm, which is indicated by a vertical blue dashed line. Below this depth, the FIB had milled beyond the Bi$_2$Se$_3$ and was milling the underlying substrate. The linearity of the mill depth vs. dose before this point suggests that there was minimal sample heating during the exposure times used.

\begin{figure*}[ht]
\centering
\includegraphics{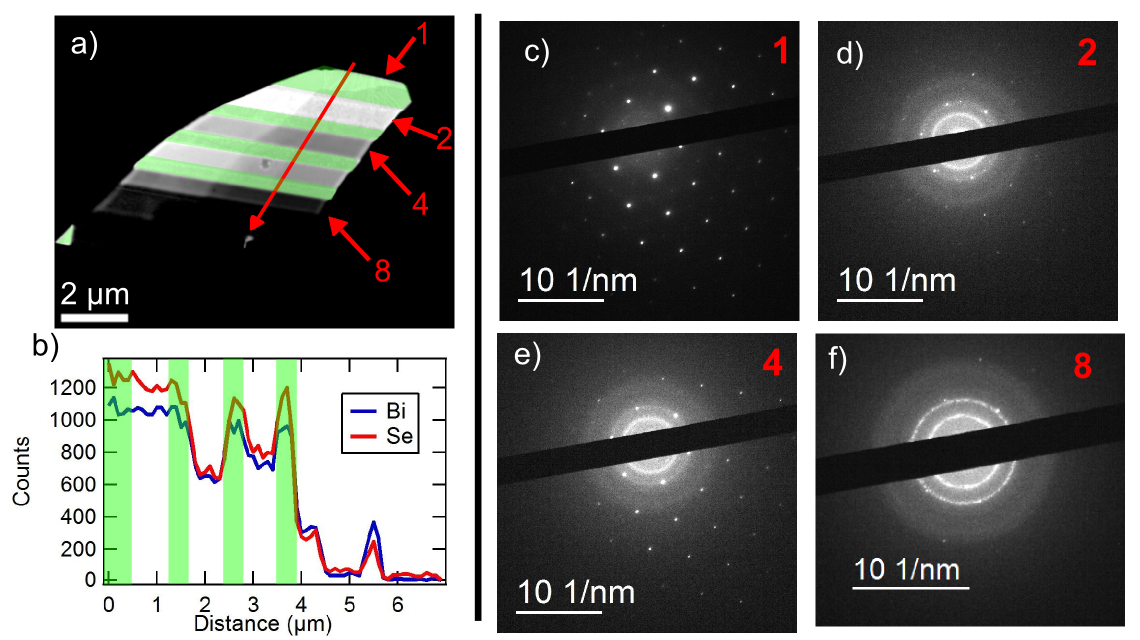}
\caption{\label{Thinned_Diff}
TEM image, EDS trace, and selected area diffraction spots from selected regions on a Bi$_{2}$Se$_{3}$ flake thinned using the focused ion beam at the 10 pA setting. SAED patterns from all regions are shown in the Supplement. (a) Counts per second image of thinned Bi$_{2}$Se$_{3}$ flake. (b) Trace of EDS intensities of Bi (blue) and Se (red) across the sample as represented with a red arrow crossing the sample in (a). (c) Diffraction from bulk region, region 1 in (a). (d) Diffraction from region thinned for 1 second, region 2 in (a). (e) Diffraction from region thinned for 5 seconds, region 4 in (a). (f) Diffraction from region thinned for 10 seconds, region 8 in (a). 
}
\end{figure*}

Fig. \ref{Thinned_Diff} presents EDS and SAED analysis of a FIB-thinned sample on a silicon nitride TEM window. Fig. \ref{Thinned_Diff}(a) shows a counts per second image of an exfoliated and milled Bi$_2$Se$_3$ flake. Regions of the flake that were not milled are highlighted in green. The total ion doses used for each region are noted in the Supplement and range from 10$^{15}$ cm$^{-2}$ to 2x10$^{16}$ cm$^{-2}$. The effect of ion milling on the flake is evident in both the brightness of the stripes in the image and in the x-ray counts in the EDS line scan (path indicated by a red line across the sample in Fig. \ref{Thinned_Diff}(a)), as shown in Fig. \ref{Thinned_Diff}(b). The green regions in Fig. \ref{Thinned_Diff}(b) again denote the unmilled regions. The first milled region, which recieved an ion dose of 10$^{15}$ cm$^{-2}$, shows minimal change in Se:Bi count ratio. Regions subjected to higher doses show a decrease in the Se:Bi ratio, which indicates preferential removal of Se. This is expected due to Se's lower mass. The decrease in the ratio is most extreme in the small amount of leftover material at a horizontal distance 5.5 $\mu$m from the start of the line scan. At this position, the Bi signal dominates. EDS signatures from Ga were observed only in the areas below region 8, which has been milled down below a thickness of 10 nm. This suggests that the Ga ions largely pass through the Bi$_2$Se$_3$ and become embedded in the silicon nitride. SAED images from several bands across the flake are shown in Fig. \ref{Thinned_Diff}(c-f). Fig. \ref{Thinned_Diff}(c) is from the top un-etched region, labeled 1 in Fig. \ref{Thinned_Diff}(a), and shows a hexagonal diffraction pattern consistent with bulk Bi$_2$Se$_3$. The SAED images from regions 2 and 4, Figs. \ref{Thinned_Diff}(d) and (e) respectively, show both the hexagonal pattern and a distinct ring with a diameter of 6.3 nm$^{-1}$. The diffraction pattern from region 8 shows both the 6.3 nm$^{-1}$ ring and some evidence of polycrystalline Bi$_2$Se$_3$. The polycrystalline structure is indicated by a ring of spots with a diameter of 9.1 nm$^{-1}$, which corresponds to a lattice spacing of 0.22 nm. The 6.3 nm$^{-1}$ diameter ring in the milled regions suggests that milling creates an amorphous surface. Given the decrease in the Se:Bi ratio, the surface is likely primarily Bi. The ring diameter equates to a lattice spacing of 0.33 nm, which means the surface could correspond to a bismuth oxide layer formed from air exposure after FIB milling\cite{Ying_Wang_2015,Juneho_In_2011} or recrystalized bismuth\cite{Jing_Chen_2007,Xiaofeng_Chang_2015}.

\begin{figure*}[ht]
\centering
\includegraphics{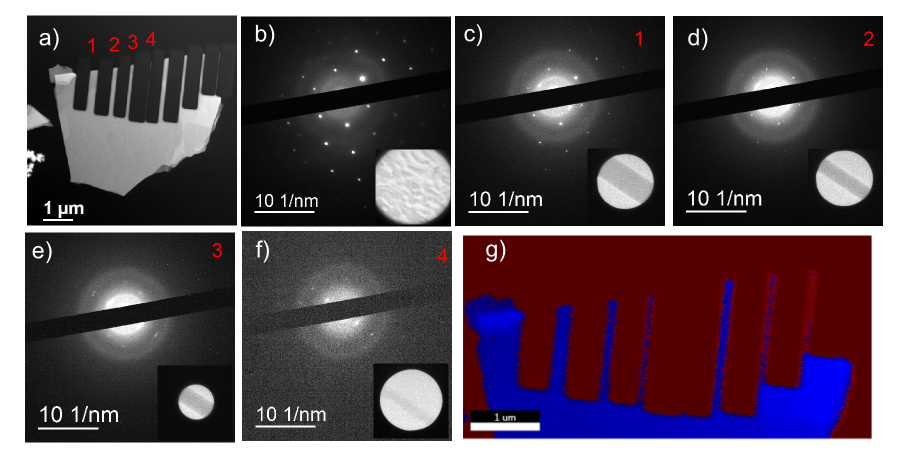}
\caption{\label{Wire_Diff}
Bi$_{2}$Se$_{3}$ sample cut into wires with diffraction and EDS imaging. (a) Counts per second image of Bi$_{2}$Se$_{3}$ wire sample.  (b) SAED pattern, and inset image, of unaltered Bi$_{2}$Se$_{3}$ region. (c) SAED of wire 1, width of 260 nm, with inset of selected region. (d) SAED of wire 2, width of 160 nm, with inset of selected region. (e) SAED of wire 3, width of 120 nm, with inset of selected region. (f) SAED of wire 4, width of 26 nm, with inset of selected region. (g) EDS map of wire sample, blue represents regions containing Bi and Se and red represents their absence.
}
\end{figure*}

Results from sculpting and milling wires from a flake are presented in Fig. \ref{Wire_Diff}. Fig. \ref{Wire_Diff}(a) shows a counts per second EDS image of a sample milled to make eight wires of varying widths, from 260 nm to 26 nm. Areas in which the beam etched entirely through the Bi$_2$Se$_3$ and into the silicon nitride appear as lighter gray regions in between the wires. The lighter color is indicative of Ga embedded in the substrate. Elemental maps of the Ga, Se, and Bi in the sample are presented in the Supplement. Numbered 1-4 in the image are wires selected for further analysis. Fig. \ref{Wire_Diff}(b) shows the diffraction pattern from the bulk Bi$_2$Se$_3$  region of the flake with the selected area inset. SAED from wires 1-4 are shown in Fig. \ref{Wire_Diff}(c-f) respectively, with the selected areas inset. As the wires decrease in width, the Bi$_2$Se$_3$ diffraction spots become more faint. For wire 3 (width 120 nm) only one pair of Bi$_2$Se$_3$ diffraction spots is visible, and for wire 4 (width 26 nm) none are visible. The destruction of the Bi$_2$Se$_3$ as a function of wire width is further evidenced by the EDS elemental mapping shown in Fig. \ref{Wire_Diff}(g). In the map, blue areas represent regions containing Bi/Se/Si, and the red areas represent Si only. The Ga signal, while observable in an elemental map, is low enough that the EDS software does not distinguish it in the phase maps. As can be seen in the map, wire 4 is not distinguishable from the substrate. This indicates that although some material remains, as shown in the counts per second image, the wire may no longer be Bi$_2$Se$_3$. This supports what was observed in the diffraction pattern in Fig. \ref{Wire_Diff}(f), which shows no evidence of a Bi$_2$Se$_3$ crystal structure. These results are comparable to the transport results in Bhattacharyya et. al.\cite{Bhattacharyya_QOsc}. In the article the authors Ahranov-Bohm oscillations, in FIB-cut nanowires, that show a period associated with wire cross-sections smaller than their physical cross-sections, indicating damage to the wires from the FIB milling. The authors suggested this was due to deformations/cracking or Ga$^{+}$ ion implantation; our results suggest that the dominant factor is defect creation, which is supported by the minimal Ga signal in the EDS of the thinned sample and by the disappearance of the Bi$_2$Se$_3$ diffraction spots as a function of width in the wires. It should also be noted that we observe no evidence of large scale cracking from the ion milling, which would appear as splitting in the SAED pattern and as visible cracks in the TEM images of the wires.

\begin{figure*}[ht]
\centering
\includegraphics{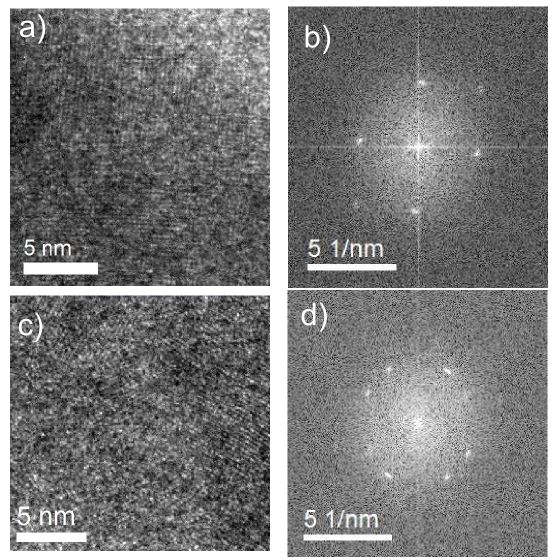}
\caption{\label{HR_FFT}
High resolution TEM and FFT images of wire edges.(a) HRTEM image of edge of wire 2 from Fig.\ref{Wire_Diff}(a). (b) FFT of (a). (c) HRTEM image of edge of wire 4 from Fig.\ref{Wire_Diff}(a). (d) FFT of (c).
}
\end{figure*}

In order to look more carefully at the material along the wire edges and the structure of wire 4, we collected high resolution TEM (HRTEM) images. These images and their fast Fourier transforms (FFT) are shown in Fig. \ref{HR_FFT}. The HRTEM image of the edge of wire 2 is shown in Fig. \ref{HR_FFT}(a) and its FFT is shown in Fig. \ref{HR_FFT}(b). On this edge, only one pair of points in the FFT is associated with the crystal structure of Bi$_2$Se$_3$, while the most prominent pairs indicate a lattice spacing of 0.3 nm.  The HRTEM and FFT images of wire 4 are shown in Fig. \ref{HR_FFT}(c) and (d) respectively. Here, there are no observable points associated with Bi$_2$Se$_3$. The dominant structure observed in the wire 4 FFT is the 0.3 nm spacing also found at the edge of wire 2. The FFT of a larger portion of wire 4 is shown in the supplement and does indicate that some of the wire may remain Bi$_2$Se$_3$, but the dominant signal is from the 0.3 nm spacing. This further supports the results from Fig. \ref{Wire_Diff}(f) and Fig. \ref{Wire_Diff}(g). The FFT and HRTEM images of the wire edges indicate that while some crystal order exists, the edges are largely not Bi$_2$Se$_3$. These images, in particular the results from Fig. \ref{HR_FFT}(a) and (b) suggest that the material at the edge has formed into a cubic structure. This material could be a bismuth oxide\cite{Ying_Wang_2015,Juneho_In_2011} or crystalline bismuth\cite{Jing_Chen_2007,Xiaofeng_Chang_2015}, similar to that suggested by the thinning results. However, the material by the edge may not directly reflect the edge itself.

\begin{figure*}[ht]
\centering
\includegraphics{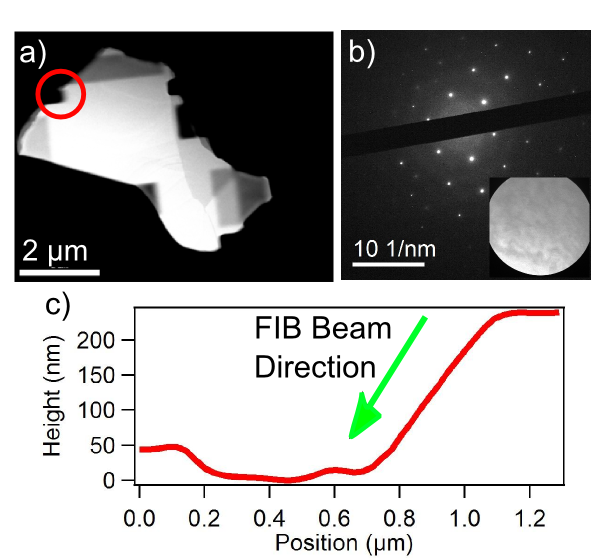}
\caption{\label{Angle_Cut}
Images and analysis of Bi$_{2}$Se$_{3}$ flake cut at a 52$^\circ$ angle from the substrate normal. (a) TEM image of angle-cut flake. (b) SAED image of region circled in red in (a). Inset is the image of the actual area selected for taking diffraction. (c) Height versus position slice from FIB cut region circled in red in (a). Green arrow indicates the approximate direction of the beam relative to the sample.
}
\end{figure*}

To study the effect of the ion beam on the edges it creates during milling, without the interference of excess material, a sample was milled at an angle. The TEM and AFM results from this sample are shown in Fig. \ref{Angle_Cut}. The red circle on the TEM image in Fig. \ref{Angle_Cut}(a) highlights the primary analyzed region, though the sample was milled in multiple locations. In this region the beam was incident from right of the sample at an angle of 52$^\circ$ from the substrate normal. Fig. \ref{Angle_Cut}(b) shows the diffraction from this region, with the selected area shown in the inset. The diffraction pattern from this area is hexagonal and has a lattice parameter of 0.21 nm. Unlike the thinned regions presented in Fig. \ref{Thinned_Diff}, no amorphous rings appear, which indicates the angle-cut surface is minimally damaged and indistinguishable from un-milled Bi$_2$Se$_3$. In an AFM slice of this region (Fig. \ref{Angle_Cut}(c)), we find that the dose from the beam (direction shown by the green arrow), was also able to mill into the silicon nitride window. The surface of the cut is smooth, and the angle of the cut relative to the substrate normal is 58$^\circ$. The higher angle of the cut indicates that some material was removed in excess.

\section*{Discussion}
In conclusion, FIB milling of Bi$_2$Se$_3$ shows promise as method for sculpting this topological insulator to create application-specific geometries. While FIB-thinning the sample creates a uniform and smooth surface, it also damages the material and creates regions that are Se-deficient and partially amorphous. The edges created when cutting wires in the Bi$_2$Se$_3$ are smooth, and their diffraction patterns are indistinguishable from bulk regions. The excess material from direct exposure is shown to be crystalline, but the exact nature of this material was not determinable in this study. The quality of the wires decreased below a width of approximately 150 nm. Smaller wires that maintain the Bi$_2$Se$_3$ crystal may be achievable, however, depending on the characteristics of the FIB tool. Overall, FIB control of Bi$_2$Se$_3$ nanostructures can potentially lead to interesting new physics and applications. 

\section*{Methods}

\subsection*{Nanostructure Transfer}
Nanostructures were mechanically exfoliated from bulk Bi$_2$Se$_3$ obtained from Alfa Aesar (99.999\% purity). For Bi$_2$Se$_3$ samples on SiO$_2$, they were directly exfoliated onto the substrate. For samples on silicon nitride membranes, the Bi$_2$Se$_3$ was exfoliated onto PMMA, then positioned and transferred onto a 100 nm thick membrane substrate using a micromanipulator and microscope method outlined in Mlack et al.\cite{Mlack_SciRep_2017}. After positioning, acetone was used to remove the PMMA and complete the transfer of the Bi$_2$Se$_3$ to the membrane. In between all measurements, samples were stored in a vacuum desiccator to reduce the effects of oxidation.

\subsection*{FIB Milling}
The samples were milled using a FEI Strata DB235 focused ion beam at 30 kV and 10 pA beam current setting.
 
\subsection*{TEM Measurement}
The samples were analyzed using a JEOL 2010F TEM operating at 200 kV.

\subsection*{AFM Measurement}
The AFM was a Brucker Icon atomic force microscope operated in tapping mode. 

\bibliography{Bi2Se3_Paper_Bibtex}

\section*{Acknowledgments}
This work was supported under by the National Science Foundation through the grant EFRI 2-DARE 1542707. Use of University of Pennsylvania Nano/Bio Interface Center instrumentation is acknowledged.

\section*{Author contributions statement}

J.T.M., S.F., and M.D. devised experiments. J.T.M. and S.F. prepared samples. J.T.M. performed FIB milling. S.F. performed AFM and TEM measurement. J.T.M. analyzed AFM and TEM data and prepared all figures and results discussion. All authors contributed to the manuscript preparation.

\section*{Additional information}
\textbf{Competing financial interests}: The author(s) declare no competing financial interests.

\section*{Figures}

\end{document}